\documentclass[twocolumn]{article}
\usepackage{amscd,amssymb,amsmath,latexsym,enumerate}
\usepackage[mathscr]{euscript}
\usepackage{mathrsfs}
\usepackage{epsfig}
\usepackage{fancybox}
\usepackage{verbatim}
\usepackage{tikz}
\usepackage{tikz-cd}
\usepackage{todonotes}
\usepackage{multicol}
\usepackage{graphicx}
\usepackage{mathtools}

\usepackage[small,nohug]{diagrams}
\diagramstyle[labelstyle=\scriptstyle]

\usepackage{color}

\usepackage{xcolor}
\definecolor{MyBlue}{cmyk}{1,0.13,0,0.63}
\definecolor{MyGreen}{cmyk}{0.91,0,0.88,0.52}
\newcommand{\mylinkcolor}{MyBlue}
\newcommand{\mycitecolor}{MyGreen}
\newcommand{\myurlcolor}{black}

\usepackage{hyperref}
\hypersetup{%
  bookmarksnumbered=true,bookmarksopen=false,%
  plainpages=false,
  linktocpage=true,%
  colorlinks=true,breaklinks=true,%
  linkcolor=\mylinkcolor,citecolor=\mycitecolor,urlcolor=\myurlcolor,%
  pdfpagelayout=OneColumn,%
  pageanchor=true,%
}

\title{Topological indices in condensed matter}

\author{Hermann Schulz-Baldes
\\
Friedrich-Alexander-Universit\"at Erlangen-N\"urnberg 
\\
Department Mathematik, Cauerstr. 11, D-91058 Erlangen, Germany                   
}

\date{ }

\newcommand{\CM}{{\mathbb C}}
\newcommand{\NM}{{\mathbb N}}
\newcommand{\RM}{{\mathbb R}}

\newcommand{\TM}{{\mathbb T}}
\newcommand{\ZM}{{\mathbb Z}}
\newcommand{\PM}{{\mathbb P}}

\newcommand{\Aa}{{\cal A}}
\newcommand{\Ee}{{\cal E}}

\newcommand{\EE}{{\bf E}}

\newcommand{\Bb}{{\cal B}}

\newcommand{\Ff}{{\cal F}}

\newcommand{\Tt}{{\cal T}}
\newcommand{\Rr}{{\cal R}}

\newcommand{\Mm}{{\cal M}}

\newcommand{\Kk}{{\cal K}}

\newcommand{\one}{{\bf 1}}

\newcommand{\Tr}{\mbox{\rm Tr}}

\newcommand{\SF}{{\rm Sf}}

\newcommand{\Ch}{{\rm Ch}} 
\newcommand{\Ind}{{\rm Ind}} 
\newcommand{\Ker}{{\rm Ker}} 
\newcommand{\Ran}{{\rm Ran}} 
\newcommand{\sgn}{{\rm sgn}} 
\newcommand{\Sig}{{\rm Sig}}

\newcommand{\GFunc}{f_{\mbox{\rm\tiny Ind}}}
\newcommand{\FFunc}{f_{\mbox{\rm\tiny Exp}}}

\newcommand{\Exp}{{\rm Exp}}

\begin{document}

\maketitle

\abstract{
This contribution describes the mathematical theory of topological indices in solid state systems composed of non-interacting Fermions. In particular, this covers the spectral localizer and the bulk-boundary correspondence.
}







\vspace{1cm}

\section{Short history}
\label{sec-History}

Let us start out this review with a brief description of the most influential works of theoretical and mathematical physics that ultimately lead to the theory and practice of topological indices in a solid state context. Such an index first appears in the 1982 work of Thouless,  Kohmoto, Nightingale, and den Nijs \cite{TKN} on integer quantum Hall systems. They consider a Hamiltonian $H$ on the tight-binding Hilbert space $\ell^2(\ZM^2)$ with constant rational magnetic flux through each unit cell, namely the so-called Harper Hamiltonian. A Bloch-Floquet (or discrete Fourier) transform $\Ff$ partially diagonalizes the Hamiltonian $\Ff H\Ff^*= \int^\oplus_{\mathbb T^2} dk \, H_k$ into a smooth matrix-valued function $k\in\TM^2\mapsto H_k$. The same holds for other smooth functions of $H$ such as the Fermi projection $P=\chi(H\leq \mu)$ for a Fermi level $\mu$ lying inside a gap of the spectrum of $H$, so that the characteristic function can be replaced by a smooth one. The authors note that there is a well-defined and integer-valued (first) Chern number
\begin{equation}
\label{eq-Ch2d}
\mathrm{Ch}(P)
\;=\;
2\pi\imath\int_{\mathbb{T}^2} \frac{dk}{(2\pi)^2} \, \Tr\big ( P_k [\partial_{k_1} P_k,\partial_{k_2} P_{k} ] \big )
\;,
\end{equation}
where $\imath  = \sqrt{-1}$, and $\partial_{k_j}$ are the partial derivatives, and the brackets denote the commutator. They also show that this integer is given by a diophantine equation involving the flux and the gap number. At first sight, it is surprising that the integral $\mathrm{Ch}(P)$ is an integer, but this fact and, in particular, the choice of the normalization factors in \eqref{eq-Ch2d}  was already well-known in the differential topology community for decades. The second main point of \cite{TKN} is that $\mathrm{Ch}(P)$ is actually equal to the Hall conductivity as given by the Kubo formula of linear response theory. Hence, the topological index $\mathrm{Ch}(P)$ is an experimentally observable quantity, responsible for the integer quantum Hall effect. It was, however, clear to the physics community that the existence of Anderson localized states near the Fermi was another crucial element in the theoretical explanation of the experiment, because the presence of these states implies the constancy of the Hall conductivity w.r.t. variations in the particle density and the magnetic field. This requires to study models with random potentials which are hence not periodic, so that the formula \eqref{eq-Ch2d} makes no sense. In a seminal work \cite{Bel}, Bellissard  showed how $\mathrm{Ch}(P)$ can be generalized to random operators in the framework of Connes' noncommutative geometry \cite{Connes94}. Taking this point of view, $\mathrm{Ch}(P)$ is a pairing of a $K_0$-group of the algebra of homogeneous operators with a suitable cyclic cocycle representing a $K$-homology class. This immediately implies stability properties of $\mathrm{Ch}(P)$.  In parallel, Avron, Seiler and Simon provided a rigorous version of Laughlin's argument in which $\mathrm{Ch}(P)$ appears as a quantized quantity with $k_1$ and $k_2$ being inserted fluxes \cite{ASS0}. The Chern number gained further importance when Haldane exhibited a gapped Hamiltonian with periodic instead of constant magnetic field and, nevertheless, a non-vanishing Chern number $\mathrm{Ch}(P)$ \cite{Hal}. The two works \cite{TKN,Hal} were an integral part of the 2016 Nobel Prize award. Roughly stated, this review is about extensions of and supplements to this approach to topological insulators. 

\vspace{.2cm}

The first important extension is to introduce other topological invariants for other dimensions $d\not= 2$, such as higher Chern numbers and higher winding numbers for Fermi projection or Fermi unitary of a chiral system \cite{RSFL}. This is also possible for systems with broken translation invariance, and merely under the hypothesis that the Fermi level $\mu$ lying in a spectral interval with Anderson localization \cite{BES,PS}. All these invariants are called ``complex'', because they require no real symmetry such as time-reversal or particle-hole symmetry (TRS or PHS). A special case is the second Chern numbers that appeared in the early mathematical contribution \cite{ASSS}. All this is described in Section~\ref{sec-Types}. Connes' work already contained {index theorems} showing  that $\mathrm{Ch}(P)$ is equal to the index of a suitably constructed Fredholm operator \cite{ASS,BES}. This generalizes to higher complex topological invariants \cite{PLB,PS}, see Section~\ref{sec-Index}.

\vspace{.2cm}

A new idea originating in the work of Kane and Mele \cite{KM2} was that also time-reversal invariant Hamiltonians with half-integer spin can be topologically non-trivial. The associated $\ZM_2$-invariant has not been linked to a directly observable quantity, but via a so-called {bulk-boundary correspondence} (BBC) it guarantees the existence of robust edge states which many believe to be relevant for the so-called quantum spin Hall effect (as this effect is stable under perturbations breaking the TRS, it has been argued \cite{SB1,DS1} that rather the more robust spin Chern numbers \cite{Pro} are at the origin of the phenomena). The concept of such so-called symmetry protected topological phases was largely extended \cite{RSFL,BHZ} to particle-hole, chiral, spatial symmetries, {\it etc.}. In all concrete models, the origin of non-trivial topology in condensed matter systems is always in band inversion terms of the underlying periodic systems.  For non-interacting Fermions and focusing only on TRS and PHS, this leads to a doubly $8$-periodic classification table \cite{Kit} of gapped topological phases for all Altland-Zirnbauer classes by the $K$-theory class of top degree (namely the $d$-dimensional Chern or winding number of the Fermi projection). There are, however, also other so-called weak invariants that do not appear in the classification table and are relevant for physical phenomena ({\it e.g.} the $3d$ quantum Hall effect \cite{KHW}). Somewhat later, an index-theoretic approach to topologically protected Fermion systems was put forward,  allowing to also deal with disordered systems in a strongly disordered phase \cite{SB5,GS}. All this is described in Section~\ref{sec-Z2Index}. Even though it will not be further discussed below, let us mention that symmetry protected phases also exist in a framework of interacting systems (composed of particles or spins), with conserved quantities such as particle number, parity, spin rotation, {\it etc.}.

\vspace{.2cm}

Parallel to the approach based on Chern numbers, winding numbers and indices of Fredholm operators described above, Loring and Hastings \cite{LH,Lor} put forward various ad-hoc suggestions for alternative indices, some inspired by the work of Kitaev (notably, Appendix~C in \cite{Kit0}). While this was initially disconnected from earlier work, a series of papers \cite{LS1,LS2} shows that suitable (and necessary) modifications lead to the notion of the {spectral localizer} which indeed has well-defined topological number (its half-signature at finite volume) that is equal to the index of Fredholm operators appearing in the index theorems of earlier works.  The spectral localizer provides a particularly efficient tool for numerical computations. It has been extended to deal with $\ZM_2$-indices in systems with real symmetries \cite{DS2} and weak invariants \cite{SS1}. This is the topic of Section~\ref{sec-SpecLoc}.

\vspace{.2cm}

Another breakthrough was Hatsugai's work showing that the Chern number \eqref{eq-Ch2d} in the Harper model determines the number of edge channels for operators restricted to a half-space \cite{Hat}. Using an exact sequence connecting bulk, half-space and edge $C^*$-algebras, it was possible to understand this equality as a manifestation of $K$-theoretic connecting maps \cite{SKR,KRS}. There were numerous more functional analytic follow-ups on this BBC, {\it e.g.} \cite{EG,GP,EGS}. It will be explained in Section~\ref{sec-Exact} how the robust concept of exact sequences can be transposed to bulk-defect situations.

\vspace{.2cm}

Initially, the theory of indices as sketched above was used to describe electrons in gapped solid state systems, namely quantum Hall systems and later on topological insulators. It was then applied and extended to many other systems, such as driven Floquet systems, non-hermitian Hamiltonians, semimetals, bosonic Bogoliubov-type Hamiltonians, topological photonics, mechanical topological systems, topological meta-materials. Hence, the theory of topological indices (together with the BBC, defect states, spectral localizer, {\it etc.}) has become an established and universal ele\-ment for the analysis of periodic wave-type equations and their perturbations. Needless to say, this is not the place to cite the vast physics literature.  This review merely offers a brief overview of the mathematical description of topological indices for non-interacting (fermionic) electrons in a condensed matter system with, moreover, a focus on the contributions of the author and his collaborators. Along the way many, but by far not all important contributions of other authors from the mathematical physics community are cited. A more extended list of reference and a more detailed expositions can be found in \cite{PS,Pro2,SSt}, for the spectral localizer in \cite{DSW}.

\section{Topological indices of covariant systems}
\label{sec-Types}

As a set-up covering many interesting situations, let us consider a bounded one-particle Hamiltonian $H=H^*$ acting on a $d$-dimensional tight-binding Hilbert space $\ell^2(\ZM^d,\CM^L)$ with $L$ degrees of freedom per site. Typically, it is a random family $H=(H_\omega)_{\omega\in\Omega}$ given by the sum of a kinetic energy involving finite range hopping and a potential energy composed of a periodic and possibly a random part. Here $\Omega$ is a compact configuration space on which is given a $\ZM^d$-action $T$ implementing space translations. The main assumption is that $H$ is covariant w.r.t. to the magnetic translation $a\in\ZM^d\mapsto U_a$ on $\ell^2(\ZM^d,\CM^L)$ associated to a constant magnetic field, namely that $U_a H_\omega U^*_a=H_{T_a\omega}$. Furthermore, there shall be given a $T$-invariant ergodic probability measure $\PM$ on $\Omega$. For a periodic system, $\Omega$ is a finite set equipped with the uniform distribution, while for an i.i.d. random (matrix) potential it contains a Tychonov product with a product measure. The formalism also applies to quasicrystals. Summing up, the above data provides a homogeneous system in the sense of Bellissard \cite{Bel}, see also \cite{PS,SSt}. The norm-closure of the set of all short-ranged homogeneous operators provides a $C^*$-algebra $\Aa_d$ that is actually a faithful representation of a crossed product $C(\Omega)\rtimes \ZM^d$ of $C(\Omega)$ with the action $\ZM^d$, possibly twisted by a constant magnetic field. This algebra is equipped with a tracial normalized state specified by the trace per unit volume:
\begin{align*}
\Tt(A)
&
\;=\;
\lim_{N\to\infty}\frac{1}{(2N+1)^d}\,\sum_{n\in [-N,N]^d}\,\Tr\;\langle n|A_\omega|n\rangle
\\
&
\;=\;
\EE_{\mathbb{P}}\; 
\Tr\;\langle 0| A_\omega|0\rangle
\;,
\end{align*}
where $A=(A_\omega)_{\omega\in\Omega}\in\Aa_d$. The convergence for $N\to\infty$ holds $\PM$-almost surely by Birkhoff's ergodic theorem (for the ergodic action of $\ZM^d$ on $\Omega$), which, due to the covariance relation, also shows that the almost sure value is given by the average over $\PM$ on the r.h.s.. Non-commutative derivations $\nabla_j$ are densely defined by
$$
\nabla_{j}A_\omega
\;=\;
\imath[X_j,A_\omega]
\;,
\qquad
j=1,\ldots,d\;,
$$
where the position operators are given by $X_j|n\rangle=n_j|n\rangle$. If $A\in\Aa_d$ is periodic with Fourier decomposition $\Ff A\Ff^*= \int^\oplus_{\mathbb T^d} dk \, A_k$, then 
\begin{align}
&\Tt(A)  \;=\;\int_{\mathbb T^d} \frac{dk}{(2\pi)^d} \, \Tr(A_k)\;,
\nonumber
\\
& \Ff \,\nabla_j A\,\Ff^*  \;=\;\int^\oplus_{\mathbb T^d} dk \,(\partial_{k_j} A)_k
\;.
\label{eq-CorrespPeriodic}
\end{align}
Hence, the tracial state $\Tt$ and the derivations $\partial_j$ are the non-commutative analogues of analysis tools in the framework of Bloch theory: integrating over and deriving w.r.t. quasi-momenta. With this in mind, one can generalize many formulas from the solid state literature to the non-commutative framework  which also covers disordered models. An example is the Chern number $\Ch(P)$ of the Fermi projection $P=\chi(H\leq \mu)$ below the Fermi level $\mu$, as given by \eqref{eq-Ch2d}. Let us directly spell out the higher-dimensional extensions. For an index set $I\subset\{1,\ldots,d\}$ of even cardinality $|I|$, the $I$-th even Chern number of $P$ is
\begin{equation}
\label{eq-EvenChern}
\mbox{\rm Ch}_{I} (P)
\;=\;
\frac{(2\imath \pi)^{\frac{|I|}{2}}}{\frac{|I|}{2}!}\;  \sum_{\rho\in S_I} (-1)^\rho \,\Tt\Big(P\prod_{j=1}^{|I|} \nabla_{\rho_j} P  \Big)
\;.
\end{equation}
Here the sum runs over bijections $\rho:\{1,\ldots,|I|\}\to I$ for which the definition of the signature $(-1)^\rho$ is extended using the natural order on $I$. For $d=2$ and $I=\{1,2\}$, this indeed reduces to \eqref{eq-Ch2d} under the correspondence \eqref{eq-CorrespPeriodic}. For $d=3$ and $I=\{1,2\}$ or $I=\{1,3\}$ or $I=\{2,3\}$, this gives the three $2d$-Chern numbers responsible for the $3d$ quantum Hall effect \cite{KHW}. Furthermore, the case $d=4$ and $I=\{1,2,3,4\}$ leads to the second Chern number \cite{ASSS} that is relevant for the quantization of the magneto-electric response \cite{PS}. If $|I|=d$, then $\mbox{\rm Ch}_{I} (P)$ is called a strong topological invariant, while otherwise one speaks of weak invariants. The definition \eqref{eq-EvenChern} only contains derivatives in the coordinate axis, but by taking linear combinations it is also possible to consider other directional derivatives. This is useful, {\it e.g.}, if one considers half-spaces that cut the lattice $\ZM^d$ in arbitrary directions \cite{SSt}. The prefactors in \eqref{eq-EvenChern} are chosen as in \cite{PS,SSt}, but only deviate slightly from Connes' choices \cite{Connes94}. In fact, in the framework of non-commutative geometry, $\mbox{\rm Ch}_{I} (P)$ is merely the pairing of a cyclic cohomology class specified by $I$ and the $K$-theory class of $P$. The choice of prefactors assure that the strong invariants are integer-valued. In general, weak invariants take real values. However, general principles \cite{Connes94} imply that these values are robust under certain perturbations and lie in a countable dense set. In the particular case $I=\emptyset$ where $\mbox{\rm Ch}_{I} (P)=\Tt(P)$ is merely the density of states in the gap, this is well-known as the gap-labelling theorem \cite{Bel}. For other $I$, the dependence of this image on the magnetic field can be computed iteratively using the generalized Streda formula for derivatives of $\mbox{\rm Ch}_{I} (P)$ w.r.t. the magnetic field  \cite{PS}. The final comment is about the domain of definition of $\mbox{\rm Ch}_{I} $. It is certainly well-defined if $\mu$ lies in a gap of $H$. Furthermore, if $\mu$ is a region of strong Anderson localization, then $\mbox{\rm Ch}_{I} (P)$ is well-defined and the strong invariant is actually constant as $\mu$ varies in a spectral interval with Anderson localized states \cite{BES,PS}. This fact is of crucial importance for the theoretical understanding of the quantum Hall effect. More recently \cite{SSt}, it has been shown by non-commutative harmonic analysis (Sobolev and Besov spaces) that $\mbox{\rm Ch}_{I} (P)$ is also well-defined for $\mu$ being precisely at the pseudo-gap of a semimetal. This makes a mathematical ana\-lysis of topological semimetals feasible.

\vspace{.2cm}

The invariants defined in \eqref{eq-EvenChern} are called even because they involve an even number of derivatives. Odd invariants have an index set $I$ of odd cardinality. For $|I|=1$, such invariants are called winding numbers, for $|I|\geq 3$ higher winding numbers. To associate such an invariant to a given Hamiltonian $H\in\Aa_d$, one has to suppose that it satisfies a so-called chiral symmetry of the form
\begin{equation}
\label{eq-ChiralSym}
J\,H\,J\;=\;-\,H\;=\;
\begin{pmatrix} 0 & A^* \\ A & 0 \end{pmatrix}
\;,
\qquad
J
\;=\;
\begin{pmatrix} \one  &  0 \\ 0 & -\one \end{pmatrix}
\;,
\end{equation}
where $L$ is even and $\one=\one_{\frac{L}{2}}$. If, moreover, $\mu=0$ lies in a gap of the spectrum of $H$, $A$ is an invertible ope\-rator,  and one can indeed expect to extract a winding number. The $I$-th odd Chern number is then defined by
\begin{equation}
\label{eq-OddChern}
\mbox{\rm Ch}_I (A)
\;=\;
\frac{\imath(\imath \pi)^\frac{|I|-1}{2}}{|I|!!}\;  \sum_{\rho\in S_{I}} (-1)^\rho \;\Tt
\Big(\prod_{j=1}^{|I|}(A^{-1} \nabla_{\rho_j}A ) \Big)
\;,
\end{equation}
where $(2n+1)!!=\prod^n_{k=1}(2k+1)$. All of the above comments transpose to these odd invariants. Best known are for $d=1$ and $I=\{1\}$ the winding numbers in the Su-Schrieffer-Heeger model, for $d=2$ and $I=\{1\}$ or $I=\{2\}$ the weak winding numbers of graphene-like semimetal models that determine the flat band of edge states \cite{SSt}, and for $d=3$ and $I=\{1,2,3\}$ the $3d$-winding number \cite{RSFL} that also appears prominently in the theory of $2d$ Floquet topological insulators \cite{SaS,ShT}, where it is applied to the evolution operator over one period (so no chiral symmetry is needed in this case). Let us also stress that an exact chiral symmetry as in \eqref{eq-ChiralSym} is not necessary, but an approximate chiral symmetry in the sense that the norm of $H+JHJ$ is small is sufficient to insure that the off-diagonal entry of $H$ is invertible and the invariant \eqref{eq-OddChern} is well-defined \cite{DS1}. This implies that both even and odd Chern numbers are defined on open sets of covariant operators. All these Chern numbers are called complex topological invariants because they do not involve a real structure (as time-reversal und particle-hole symmetries). Real topological invariants are sometimes finite groups of torsion type and can best be studied using the Fredholm operators that appear in the index theorems. These torsion invariants are described in Section~\ref{sec-Z2Index}.

\section{Index theorems}
\label{sec-Index}

The Chern numbers \eqref{eq-EvenChern} and \eqref{eq-OddChern} are non-commutative generalizations of invariants of differential topological nature associated to vector bundles. A successful and particularly influential part of 20th century mathematics connects such topological indices to analytical indices, namely to indices of suitably constructed Fredholm operators (Noether's index theorem for winding numbers, the Atiyah-Singer index theorem and the Boutet-de-Monvel index theorem). This theme has been greatly generalized into a non-commutative framework by the works of Connes, Kasparov, and many others \cite{Connes94}. The proofs of such index theorems usually involve algebraic manipulations with a geometric flavor. For the Chern number $\Ch_{\{1,2\}}(P)$ in $d=2$ this is based on Connes' triangle equality \cite{Connes94,ASS,BES}. For strong even invariants it was genera\-lized in \cite{PLB}, for strong odd invariants in \cite{PS}, and for weak invariants in \cite{PS2}. Let us describe the outcome. One builds a Dirac operator $D=\sum_{i\in I}X_i\gamma_i$ where $\gamma_1,\ldots,\gamma_d$ is an irreducible selfadjoint representation of the Clifford algebra, namely these are matrices with an anti-commutator satisfying $\{\gamma_i,\gamma_j\}=2\delta_{i,j}$. For $|I|$ even, there is some selfadjoint unitary matrix $\gamma_0$ anti-commuting with the representation. In a suitable re\-presentation, $\gamma_0$ is the third Pauli matrix and $D$ is then off-diagonal with lower left diagonal entry $D_0$. Setting $F=D_0|D_0|^{-1}$ (with an arbitrary modification on the finite-dimensional kernel of $D_0$), one can then show that $P FP$ is (semifinite) $\Tt$-Fredholm operator on the range of $P$ and its index satisfies
\begin{equation}
\label{eq-EvenChernIndex}
\mbox{\rm Ch}_{I} (P)
\;=\;
\Tt\mbox{-}\Ind_{(P\cdot P)}(P FP)
\;,
\end{equation}
whenever the l.h.s. is well-defined. If $|I|=d$ so that $\mbox{\rm Ch}_{I} (P)$ is the strong invariant, $T=PFP$ is a Fredholm operator on $\Ran(P)$ in the conventional sense and the index is $\Ind(T)=\dim(\Ker(T))-\dim(\Ker(T^*))$. Otherwise, one has $\Tt\mbox{-}\Ind_{(P\cdot P)}(T)=\Tt(\Ker(T))-\Tt(\Ker(T^*))$ (for a detailed elementary exposition of this semi-finite index theory, see \cite{DSW}). For odd cardinality of $I$, introduce the Hardy projection $E=\frac{1}{2}(D|D|^{-1}+\one)$, then $EAE$ is a Fredholm operator on $\Ran(E)$ and
\begin{equation}
\label{eq-OddChernIndex}
\mbox{\rm Ch}_{I} (A)
\;=\;
\Tt\mbox{-}\Ind_{(E\cdot E)}(EAE)
\;.
\end{equation}
Note that in the case of strong invariants, the indices are integers. The two index theorems justify the normalizations in \eqref{eq-EvenChern} and \eqref{eq-OddChern}, another justification is that the BBC takes a particularly simple form (see Section~\ref{sec-Exact}). In the case of $|I|=d$, namely the strong topological invariants, one can show that the indices on the r.h.s. of \eqref{eq-EvenChernIndex} and \eqref{eq-OddChernIndex} are almost surely constant \cite{BES,PS}. In part of the mathematical physics literature, this index for a given, fixed configuration is considered as the fundamental object. Indeed, it is possible to verify the Fredholm property of the operators on the r.h.s. of \eqref{eq-EvenChernIndex} and \eqref{eq-OddChernIndex} by merely assuming the locality of the Hamiltonian and a spectral gap ({\it e.g.} \cite{ShT,CKS}). This may be useful to just distinguish topological phases by the index, but the cohomological formula on the l.h.s. is typically related to observable quantities (such as the Hall conductivity in the case \eqref{eq-Ch2d} in the quantum Hall effect, or the chiral polarization in $d=1$, the magneto-electric effect, {\it etc.} \cite{PS}). Nevertheless, the next two sections are about strong invariants associated to fixed Hamiltonians (instead of covariant families) and given by the indices of Fredholm operators as in \eqref{eq-EvenChernIndex} and \eqref{eq-OddChernIndex}.

\section{$\ZM_2$-invariants}
\label{sec-Z2Index}

Since the pioneering work of Kane and Mele \cite{KM2}, it is appreciated that topological invariants do not necessarily have to be integer-valued and of the types described in Section~\ref{sec-Types}. More precisely, the work \cite{KM2} considers two-dimensional periodic gapped Hamiltonians with a time-reversal symmetry (TRS) and half-integer spin, and then associates to them an invariant in $\ZM_2=\{0,1\}$ which is stable under homotopies respecting the TRS. The construction of this so-called symmetry protected invariant is fairly intricate in \cite{KM2}, but there are by now numerous alternatives which all merely extract the torsion part of the even real $K$-group on the two-torus. Soon after the work \cite{KM2}, it was realized that also other symmetries can protect topological phases \cite{RSFL,Kit}. An important example is the particle-hole symmetry (PHS) which a Bogoliubov-de Gennes Hamiltonian inherits directly from the choice of representation (hence the PHS is {\it not} an extra symmetry related to some invariance of the physical system). Both the TRS and PHS involve a complex conjugation and hence a real structure on the Hilbert space, and moreover both symmetries can be either even or odd, namely the real unitary operator implementing the symmetry can either square to the identity $\one$ or to $-\one$. Explicitly, the symmetries are given by
\begin{equation}
\label{eq-SymClass}
I^*\,\overline{H}\,I\;=\;H\;\;\;\mbox{(TRS)}\;,
\qquad
K^*\,\overline{H}\,K\;=\;-\,H\;\;\;\mbox{(PHS)}\;,
\end{equation}
where $\overline{H}$ is the complex conjugation of $H$, and $I=\overline{I}=(I^*)^{-1}$ satisfies either $I^2=\one$ or $I^2=-\one$, and simi\-larly for $K$. As these  symmetries can also be present simultaneously (by combining them, the Hamiltonian then also has an induced chiral symmetry), one has a total of $8$ real symmetry classes. Together with the two complex classes (chiral symmetric systems and systems without symmetry), they make up the $10$ Altland-Zirnbauer symmetry classes. Within each class and for every fixed dimension $d$, one can now classify insulators via their Fermi projection $P$. This can be done by so-called $KR$-theory of the algebra $\Aa_d$ where the $R$ indicates that the real symmetries \eqref{eq-SymClass}  are supposed to be satisfied (note that for the PHS this means $K^*\overline{P}K=\one-P$). For periodic systems without magnetic field, this reduces to the $KR$ groups of the Brillouin torus $\TM^d$ (or the continuous functions thereon). These groups are all known and contain summands given by even and odd (winding) Chern numbers (from $\ZM$) and torsion terms with values in $\ZM_2$ \cite{Kit}. The strong invariant can also take values in $\ZM$ or $\ZM_2$, or be trivial. Within each symmetry class, these strong invariants are $8$-periodic due to Bott periodicity, and if the $8$ real symmetry classes are placed in the right order (associated to the Clifford algebras implementing the symmetries) there is also an $8$-periodicity in the class for a fixed dimension \cite{Kit,SCR,Thi,GS}. In this manner one obtains a doubly-periodic table of $64$ (strong) topological insulator classes.

\vspace{.2cm}

For  the integer-valued entries in this table, one can use the formulas \eqref{eq-EvenChern} and \eqref{eq-OddChern} even if the insulator is disordered. For the cases with $\ZM_2$-valued torsion terms, the explicit construction of the invariant in a fiber bundle framework is quite involved ({\it e.g.} as \cite{KM2}, or in \cite{ASV,GP}) and a simple cohomological formula is not available \cite{Kel,DS1}. However, the operators appearing on the r.h.s. of  \eqref{eq-EvenChernIndex} and \eqref{eq-OddChernIndex} still have a Fredholm property (in the case $I=\{1,\ldots,d\}$ of strong invariants). Because $P$ and $A$ have a symmetry property inherited from \eqref{eq-SymClass}, and also the Dirac operator has a real symmetry depending on the dimension $d$, the Fredholm operators have a symmetry property which in the case of a $\ZM_2$-entry in the periodic table implies that the index vanishes \cite{GS}. It is then, however, possible to show that the parity of the nullity is a well-defined homotopy invariant that can be used as a phase label even in the Anderson localized phase \cite{GS}. For example, for the $d=2$ and odd TRS considered in \cite{KM2}, the invariant is
\begin{equation}
\label{eq-TRSZ2}
\Ind_2(PFP)\,=\,\dim(\Ker(PFP))\,\mbox{\rm mod}\,2\;\in\;\ZM_2=\{0,1\}
\;.
\end{equation}
These phase labels are not directly equal to a physical quantity other than the integer-valued invariants (like the Chern number in \eqref{eq-Ch2d} is equal to the Hall conductivity, up to a constant). But a non-trivial $\ZM_2$-index is relevant for physical phenomena, like the existence of conducting edge modes in the quantum spin Hall effect \cite{KM2,ASV,GP} or the existence of Majorana bound states at half-flux perturbations in $p+ip$ superconductors \cite{DS}. Finally let us comment that, if the system has a supplementary (possibly only approximate) symmetry, the $\ZM_2$-invariant may be given as the parity of an integer-valued invariant. For example, the parity of the spin Chern numbers \cite{Pro} gives the $\ZM_2$-valued Kane-Mele invariant \cite{SB5}. A systematic treatment of such approximate symmetries is provided in \cite{DS1}. The associated integer-valued invariants are of the type \eqref{eq-EvenChern} and \eqref{eq-OddChern}, and thus do not pend on the real symmetries any more, and are therefore much more robust than the $\ZM_2$-invariants themselves. 

\section{Spectral localizer}
\label{sec-SpecLoc}

To motivate the definition of the spectral localizer, let us focus on a gapped Hamiltonian $H$ in dimension $d=2$. If $H$ is periodic, one then has an associated Chern number  \eqref{eq-Ch2d} which for random perturbations takes the form \eqref{eq-EvenChern}. The index theorem \eqref{eq-EvenChernIndex} shows that this topological invariant can be computed from the Fermi projection $P$ or equivalently the flat band Hamiltonian $\one-2P$ and the two components $X_1$ and $X_2$ of the position operator. These are $3$ selfadjoint operators, and the commutators are bounded. Replacing $\one-2P$ by $H$ and multiplying the positions with a small constant $\kappa>0$, one obtains $3$ selfadjoint operators $H$, $\kappa X_1$ and $\kappa X_2$ which have small commutators of order $\kappa$ (their size is measured in norm, and, of course, the commutator $[\kappa X_1,\kappa X_2]$ even vanishes). It is useful to think of these operators as non-commutative coordinates in a three-dimensional space. From these coordinates and the Pauli matrices $\sigma_1$, $\sigma_2$ and $\sigma_3$ one can build the selfadjoint operator $L_\kappa$, called the spectral localizer, and compute its square
\begin{align*}
L_\kappa & 
\;=\;
\kappa \,X_1\otimes\sigma_1+\kappa \,X_2\otimes \sigma_2-H\otimes\sigma_3
\;,
\\
(L_\kappa)^2 &
\;=\;
\big(\kappa^2 (X_1)^2+\kappa^2( X_2)^2+H^2 \big)\otimes\one
\\
& \;\;\;\;\;\;\;
+\kappa\, \imath[X_1,H]\otimes \sigma_2-\kappa \,\imath[X_2,H]\otimes\sigma_1
\;.
\end{align*}
Due to the gap of $H$ and for $\kappa$ sufficiently small, the operator $(L_\kappa)^2$ is bounded from below so that $L_\kappa$ has a gap $g$ around $0$. Moreover, $L_\kappa$ clearly has a compact resolvent, and one can check that the associated semigroup has sufficient decay to assure the existence of the $\eta$-invariant \cite{DSW}. Roughly stated, this invariant is the spectral asymmetry given by the difference of the number of positive and negative eigenvalues (of course, both of these numbers are infinite, but the definition of the $\eta$-invariant is a suitable regularization procedure to make this well-defined). For a finite selfadjoint matrix, the $\eta$-invariant simply reduces to the signature of the matrix. It is now possible to show that finite volume restrictions $L_{\kappa,\rho}$ of $L_\kappa$ to either balls $\Ran(X_1^2+X_2^2\leq \rho^2)$ or squares $\Ran(X_1^2\leq \rho^2)\cap\Ran(X_2^2\leq \rho^2)$ actually already have a spectral gap (here $H$ has simply Dirichlet boundary conditions). This assures a well-defined signature that determines the strong index pairing and therefore, due to \eqref{eq-EvenChernIndex},  also the strong Chern number. More precisely, suppose that $H$ is short range with gap $g=\|H^{-1}\|^{-1}$. If
\begin{equation}
\label{eq-HypBound}
\kappa\;\leq\;\frac{g^{3}}{12\,\|H\|\,\|[X_1+\imath X_2,H]\|}
\qquad\mbox{and}\qquad \frac{2g}{\kappa}
\;\leq\;\rho
\;,
\end{equation}
then it is shown in \cite{LS2} that $(L_{\kappa,\rho})^2\geq \frac{g^2}{4}$ and 
\begin{equation}
\label{eq-SigInd}
\Ind\big(PFP|_{\Ran(P)}\big)
\;=\;
\frac{1}{2}\;\Sig(L_{\kappa,\rho})
\;.
\end{equation}
This equality provides a very efficient tool for numerical computations of the strong topological invariants. Indeed, $L_{\kappa,\rho}$ is easy to build and the computation of the signature merely requires an LDL decomposition. Numerical results indicate that \eqref{eq-SigInd} also holds in an Anderson localization regime at the Fermi level (when $H$ has no gap), but no proof is available.  As shown in \cite{LS2}, the equality \eqref{eq-SigInd} is actually merely the numerical output of a $K$-theoretic statement: the joint spectra of the $3$ selfadjoint operators $H$, $\kappa X_1$ and $\kappa X_2$ avoid the origin and thus, after normalization, these operators form a fuzzy sphere; such fuzzy spheres are distinguished by the $K$-theory classes of the finite volume localizer (lying in the algebra of compact operators). Considerably easier proofs of \eqref{eq-SigInd} based on spectral flow have appeared afterwards \cite{LSS,SS1,DSW}. All proofs show that the equality \eqref{eq-SigInd} actually holds under the same condition in any even dimension $d$ for 
$$
L_\kappa
\;=\;
\kappa\, D\,-\,H\otimes\gamma_{0}
\;,
\qquad
\mbox{\rm where }\;D\gamma_{0}\,=\,-\gamma_{0}D
\;.
$$
This formula shows that the spectral localizer essentially consists of placing the Hamiltonian in a linear trap given by the dual Dirac operator. As such, it is centered at some point, here the origin. But one can also simply shift $D$ by $x\in\RM^d$ to $D_x=\gamma \cdot (X-x)$ and then compute the spectral localizer $L_\kappa(x)$ and its half-signature as a local index. This allows to detect topological phase boundaries by looking at the spectral flow of $L_\kappa(x)$ as $x$ varies \cite{CKS}. 

\vspace{.2cm}

Several further extensions and variations of spectral localizer techniques have been developed.  First of all, the strong odd invariants \eqref{eq-OddChernIndex} associated to a gapped chiral Hamiltonian in the form \eqref{eq-ChiralSym} are equal to the half-signature of the finite volume approximation of the odd spectral localizer given by
$$
L_\kappa
\;=\;
\kappa\,D\otimes J\,+\,H
\;,
\qquad
\mbox{\rm where }\;J\,H\,J\,=\,-H
\;,
$$ 
again provided that the equivalent of \eqref{eq-HypBound} holds \cite{LS1,DSW}. Second of all, weak invariants can be computed by using semifinite traces to compute the (typically not integer-valued) signature of a spectral localizer containing a Dirac operator that only involves the directions corresponding to the derivatives in the given weak invariant \cite{SSt}. Third of all, invariants associated to approximate symmetries (such as the spin Chern number) can be computed directly by a modified spectral localizer \cite{DS1}. Moreover, $\ZM_2$-invariants can be accessed from skew-adjoint versions of the spectral localizer by extracting the sign of its Pfaffian or possibly the sign of the determinant of its off-diagonal entry \cite{DS2}. While this reference provides a construction of the skew localizer in all $16$ relevant cases in Table~\ref{tab-class}, let us spell out the outcome in the most studied cases of Hamiltonian with an odd TRS $I^*\overline{H}I=H$ in dimension $d=2$ and $d=3$. In the two-dimensional case this corresponds  to the quantum spin Hall effect with $\ZM_2$-index already given in \eqref{eq-TRSZ2} above. Introduce the two real matrices $\Re (H)=\tfrac{1}{2}(H+\overline{H})$ and $\Im (H)=\tfrac{1}{2i}(H-\overline{H})$ which are symmetric and skew-symmetric respectively. Then the skew localizer is defined by
$$
L_\kappa
\;=\;
\begin{pmatrix} 
\Im (H) +\kappa X_1 I & \Re (H)I+\kappa X_2 \\ I\,\Re (H)-\kappa X_2 & \Im (H) -\kappa X_1 I
\end{pmatrix}
\;.
$$
It satisfies $L_\kappa=\overline{L_\kappa}=-\,(L_\kappa)^*$, relations which also hold for $L_{\kappa,\rho}$. Again under the conditions \eqref{eq-HypBound}, one then has
$$
\Ind_2\big(PFP\big)
\;=\;
\sgn(\mbox{\rm Pf}(L_{\kappa,\rho}))
\;.
$$
For $d=3$ this concerns the $\ZM_2$-topological insulators. The skew-localizer (real and skew) turns out to be off-diagonal and then the computation of the Pfaffian reduces to the determinant of its off-diagonal entry which is given by $\Rr^*(\imath\kappa D+H)\Rr$ for a suitable matrix $\Rr$ and where $D=X_1\sigma_1+X_2\sigma_2+X_3\sigma_3$. Then the $\ZM_2$-invariant is given by
$$
\Ind_2\big(E(\one-2P)E+\one-E\big)
\;=\;
\sgn\big(\det( \imath\kappa D+H)\big)
\;.
$$
It is not obvious that $\det( \imath\kappa D+H)$ is real, but this is part of the claim.

\vspace{.2cm}

Also for non-hermitian line-gapped systems, a suitable modification of the definition leads to a line-gapped spectral localizer, of which a signature can be computed by counting the number of eigenvalues to the left and right of that line. Again, this signature allows to access the topological invariants \cite{CKS}. Finally the spectral localizer can also be used for spectral analysis: for an ideal semimetal which only has a finite number of (Dirac or Weyl) points at the Fermi level, their number is equal to the approximate kernel of the spectral localizer \cite{SS2}. One can also measure the size of the Fermi surface in a metal by looking at the spectral density of the localizer at the origin \cite{Gru}. Further applications to corner or defect states are currently analyzed.

\section{Connecting indices via exact sequences}
\label{sec-Exact}

$K$-theory can be used to distinguish and classify topological cases by the class of their Fermi projection, and index theory associates numerical values to these classes. There is another important aspect of $K$-theory though: the $K$-groups of the $C^*$-algebras in an exact sequence are connected by the boundary maps, namely the index and exponential maps. There are also corresponding dual maps on the level of (cyclic) cohomology, and together this allows to connect the invariants. Alternatively, the exact sequence can be viewed as elements of $KK$-groups, and then the Kasparov product connects the invariants.  In applications to solid state systems, the so-called bulk $C^*$-algebra $\Aa_d$ mentioned in Section~\ref{sec-Types} is one of the algebras in the exact sequence. The others are associates to defects or modifications of bulk which one would like to study. The most prominent (and actually historically also the first \cite{SKR,KRS}) example of such an exact sequence is at the root of the bulk-boundary correspondence (BBC):
\begin{equation}
\label{eq-ExactSeq}
\begin{diagram}
0 &\rTo &\mathcal E_d   &\rTo{i}  &\widehat{\mathcal A}_d  &\rTo{ \mathrm{ev}}  &\mathcal A_d &\rTo &0
\end{diagram}
\end{equation}
Here $\Ee_d$ is the algebra of operators on a half-space tight-binding Hilbert space $\ell^2(\ZM^{d-1}\times\NM,\CM^L)$ which are covariant along the boundary and fall off away from the boundary, and $\widehat{\mathcal A}_d$ is the algebra of operators obtained as sums operators in $\Ee_d$ and restrictions of covariant operators from $\Aa_d$ to the half-space Hilbert space. Furthermore, $i$ is the inclusion and "ev" extracts the covariant part from $\widehat{\Aa}_d$.  As an  exact sequence of vector spaces, this sequence is split (so essentially trivial), but {\it not} as a sequence of algebras. Actually \cite{KRS,PS}, it is isomorphic to the Toeplitz extension of Pimsner and Voiculescu \cite{PV}. Also the case $d=1$ is of interest: then $\Ee_d$ is simply the algebra $\Kk$ of compact operators on $\ell^2(\NM)\otimes\CM^L$, and $\widehat{\mathcal A}_d$ is the classical Toeplitz algebra (enlarged by disorder). For larger $d$, one can show that $\Ee_d$ is isomorphic to $\Aa_{d-1}\otimes\Kk$. In particular, the complex $K$-groups satisfy $K_j(\Ee_d)=K_j(\Aa_{d-1})$, $j=0,1$. Furthermore, the definition of the even and odd Chern numbers \eqref{eq-EvenChern} and \eqref{eq-OddChern} extends to respectively projections and invertibles in the edge algebra if the trace in \eqref{eq-CorrespPeriodic} contains a trace over the compact operators $\Kk$, provided these operators satisfy a traceclass condition. How to obtain such projections and invertibles of physical relevance is discussed next.

\vspace{.2cm}

Associated to any exact sequence of $C^*$-algebras, there is always an exact sequence of $K$-groups which here is
\begin{equation}
\label{diag-6K}
\begin{diagram}
K_0(\Ee_d) &\rTo^{\ \ i_\ast \ \ } & K_0(\widehat \Aa_d) & \rTo{\ \ \mathrm{ev}_\ast \ \ } & K_0(\Aa_d) \\
\uTo{\mathrm{Ind}}                 &               \     \                        &     \     \                                 &       \     \                   &\dTo{\mathrm{Exp}} \\
K_1(\Aa_d)        & \lTo{\ \ \mathrm{ev}_\ast \ \ } & K_1(\widehat \Aa_d) & \lTo{\ \ i_\ast \ \ }    & K_1(\Ee_d)
\;.
\end{diagram}
\end{equation}
The connecting maps $\Exp$ and $\Ind$ can be computed explicitly for the cases of physical importance and connect $K$-theoretic invariants of the bulk and the boundary \cite{KRS,PS}. Suppose that  the Hamiltonian $H\in\Aa_d$ has an open gap $\Delta\subset\RM$. Then the Fermi projection $P=\chi(H\leq\mu)$ lies in $\Aa_d$ and specifies an element in $K_0(\Aa_d)$. Its image and the exponential map lies in $K_1(\Ee_d)$ and is given in terms of a half-space restriction $\widehat{H}\in \widehat{\Aa}_d$ of $H$, an arbitrary lift, by the formula
\begin{equation}
\label{ExpMapFormula}
\mbox{\rm Exp}([P]_0)
\;=\;
\Big[\exp(2\pi\imath\, \FFunc(\widehat{H}))\Big]_1
\;,
\end{equation}
where $\FFunc :\RM\to [0,1]$ is a non-decreasing smooth function equal to $0$ below $\Delta$ and to $1$ above $\Delta$. If $H$ has a chiral symmetry \eqref{eq-ChiralSym}, then the Fermi projection satisfies $JPJ=\one-P$ and specifies an element $[U]_1$ in $K_1(\Aa_d)$, and one has
\begin{align}
\mbox{\rm Ind} \big ([U]_1 \big )
\;=\;&
\Big[
e^{-\imath\frac{\pi}{2} \GFunc (\widehat{H})}
\,
{\rm diag}(\one_L,0_L)
\,e^{\imath\frac{\pi}{2} \GFunc (\widehat{H})}
\Big]_0
\nonumber
\\
&
\;-\;
\Big[{\rm diag}(\one_L, 0_L)
\Big]_0
\;,
\label{IndMapFormula}
\end{align}
where $\GFunc :\RM\to [-1,1]$ is a non-decreasing smooth odd function equal to $\pm 1$ above/below $\Delta$. 
%
%
It is remarkable that the r.h.s. in \eqref{ExpMapFormula} and \eqref{IndMapFormula} indeed define classes in $\Ee_d$, even though they are obtained by functional calculus of $\widehat{H}$ which lies in $\widehat{\Aa}_d$ and not in $\Ee_d$. The smoothness of $\FFunc$ and $\GFunc$ allow to show that the operators on the r.h.s. of \eqref{ExpMapFormula} and \eqref{IndMapFormula} satisfy the traceclass condition, and then the duality theory of pairings (see \cite{KRS,PS,SSt}) proves that for $I\subset\{1,\ldots,d-1\}$ of even and odd cardinality respectively,
\begin{align}
&
\mbox{\rm Ch}_{I\cup \{d\}}(U)\;=\;\mbox{\rm Ch}_I(\Ind(U)) 
\;,
\qquad 
\nonumber
\\
&
\mbox{\rm Ch}_{I\cup \{d\}}(P)\;=\;\mbox{\rm Ch}_I(\mbox{\rm Exp}(P))  
\;.
\label{eq-BBC}
\end{align}
Here $\Ind(U)$ and $\mbox{\rm Exp}(P)$ are representatives on the r.h.s. of \eqref{ExpMapFormula} and \eqref{IndMapFormula},   which assure that the required traceclass conditions hold. The equalities are called the BBC. For $d=2$ and $I=\{1\}$, the second equality is the celebrated BBC for the quantum Hall effect. It can then be shown that the r.h.s. $\mbox{\rm Ch}_{\{1\}}(\mbox{\rm Exp}(P))$ is actually the current density along the boundary of the sample \cite{SKR,KRS,PS}. For $d=3$ and $I=\{1,2\}$, $\mbox{\rm Ch}_{\{1,2\}}(\Ind(U)) $ is the quantized Hall conductance of the surface states \cite{PS}. The equalities \eqref{eq-BBC} can also be proved by $KK$-theory \cite{BCR,AMZ}. This explores that exact sequences themselves specify certain $KK$-group elements, and then \eqref{eq-BBC} can be seen as the outcome of various Kasparov products. It is also possible to implement real structures (as discussed in Section~\ref{sec-Z2Index}) \cite{BKR,AMZ}, but the lack of cohomological expressions for the $\ZM_2$-invariants restricts the physical applications. However, non-trivial $\ZM_2$ indices do imply the existence of boundary states in the quantum spin Hall systems (two-dimensional systems with odd TRS) \cite{KM2,ASV,GP}.

\vspace{.2cm}

The use of exact sequences in the context of solid state systems is a robust concept which is likely not yet fully exploited. There are, however, several phenomena (previously known in the physics literature) that can be understood as the manifestation of exact sequences different from  \eqref{eq-ExactSeq}. This then implies further natural stability properties. For example, in $2d$ Floquet systems, the surface modes can be understood in this manner \cite{SaS}; for $3d$ topological insulators the BBC for the magneto-electric effect is based on some exact sequence \cite{LP}; screw dislocations are also rooted in a suitable exact sequence of algebras \cite{Kub}; certain corner states are connected to quarter-plane extensions \cite{Hay}; and also scattering formulation of insulator invariants is based on an exact sequence \cite{STo}. Let us here describe yet another case that is not spelled out in detail in the literature, namely the spectral flow associated to the piercing of a monopole through the solid state system \cite{CS}. It is given by a particular path $\alpha\in[0,1]\mapsto H_\alpha$ of Hamiltonians explicitly constructed in \cite{CS}. In dimension $2$, this is merely the insertion of a flux-tube through one fixed lattice cell and, hence, has an abelian gauge field (see also \cite{DS}), but for $d\geq 3$ this is a truly non-abelian monopole (in $d=3$ a Wu-Young monopole, in $d=4$ a Dirac monopole). As the gauge can be chosen such that $H_\alpha-H_0$ is a compact operator, there is an associated spectral flow through each given and fixed gap of $H=H_0\in\Aa_d$. In even dimension, this spectral flow is precisely given by the strong topological invariant:
\begin{equation}
\label{eq-FluxSF}
\SF(\alpha\in[0,1]\mapsto H_\alpha\;\;\mbox{through }\mu)
\;=\;
\Ch_{\{1,\ldots,d\}}(P)
\;.
\end{equation}
This equality is again a manifestation of an exact sequence. Let $T(\Aa_d)$ be the trivial extension of $\Aa_d$ by the compact operators $\Kk$ on $\ell^2(\ZM^d)$ with finite-dimensional fibers to accommodate the matrix and Clifford degrees of freedom. One then has a split exact sequence $0\to \Kk \to T(\Aa_d)\to\Aa_d\to 0$, see the Appendix of \cite{DS}. If $F=D_0|D_0|^{-1}$ is the Dirac phase associated to the Dirac operator as in Section~\ref{sec-Index}, then one has $H_1=F^*H_0F$, and therefore $\alpha\in[0,1]\mapsto H_\alpha$ is an element of the mapping cone
\begin{align*}
\Mm_F
\;=\;
\Big\{&
\alpha\in[0,1]\mapsto A_\alpha\in T(\Aa_d)\;:\;A_1=F^*A_0F\;
\\
& \mbox{ and }\;A_\alpha-A_0\in\Kk
\Big\}
\;.
\end{align*}
This mapping cone lies in an exact sequence
\begin{diagram}
0 &\rTo &S\mathcal K   &\rTo{i}  &\Mm_F  &\rTo{ \mathrm{ev}}  &\mathcal A_d &\rTo &0
\end{diagram}
where $S\Kk$ is the suspension of $\Kk$ and now ev$(\alpha\in[0,1]\mapsto A_\alpha)=A_0$. The exponential map of the associated exact sequence of $K$-groups $\Exp:K_0(\Aa_d)\to K_1(S\Kk)$ applied to the Fermi projection $P=\chi(H\leq \mu)$ of the insulator $H$ is
$$
\Exp([P]_0)
\;=\;
\big[\alpha\in[0,1]\mapsto 
\exp(2\pi\imath \FFunc(H_\alpha))
\big]_1
\;,
$$
because $\alpha\in[0,1]\mapsto \FFunc(H_\alpha)$ is a lift of $P$ into $\Mm_F$ (another lift, used in \cite{DS0}, is given by $\alpha\in[0,1]\mapsto P+\alpha F^*[P,F]$). Note that $\exp(2\pi\imath \FFunc(H_\alpha))$ is a unitary operators that indeed lies in the unitization of $\Kk^\sim$ of $\Kk$ because $\exp(2\pi\imath P)=\one$ and $H_\alpha-H_0$ is compact. Furthermore, for $\alpha=1$,  one also has $\exp(2\pi\imath \FFunc(H_1))=\exp(2\pi\imath F^*\FFunc(H_0)F)=\one$ so that the path is closed. Now $K_1(S\Kk)\cong K_0(\Kk)=\ZM$ and this integer is given by the spectral flow of the path of unitaries in $\Kk^\sim$ through $-1$. Hence by Theorem~3.4 of \cite{DS0} one has
\begin{align*}
&
\Ch_{\{1,\ldots,d\}}(P)
\\
&
\;=\;
\SF\big(
\alpha\in[0,1]\mapsto 
\exp(2\pi\imath \FFunc(H_\alpha) )\;\;\mbox{through }-1
\big)
\;.
\end{align*}
Now the r.h.s. is indeed one way to compute the spectral flow in \eqref{eq-FluxSF}. In conclusion, \eqref{eq-FluxSF} is indeed routed in the mapping cone exact sequence. Modifications also allow to deal with the case of odd dimensions $d$.

\section{Delocalization of boundary states}
\label{sec-deloc}

In the physics community, the existence of conducting boundary states is often used as the defining property of topological insulators. In a mathematical description, one may rather define an topological insulator as an insulator which has some non-vanishing (strong or weak) bulk invariants. Then the BBC \eqref{eq-BBC} implies that the boundary invariants are non-trivial. In some situations, one can then prove that there are conducting boundary states. Indeed, in dimension $d=2$ and a half-space $\Ran(X_2\geq 0)$, one can express the boundary invariant as a current along the boundary \cite{SKR,KRS,PS}, namely
$$
\Ch_{\{1,2\}}(P)
=\,
\EE_\PM\,\sum_{n_2\geq 0}\,\Tr\big(\langle 0,n_2|g(\widehat{H})\,\imath [X_1,\widehat{H}]|0,n_2\rangle\big)
,
$$
where $g$ is a positive function of unit integral and support in a bulk gap. Hence the r.h.s. is the current density along the boundary and the formula shows the quantization of the boundary currents. Similarly, for a chiral insulator in dimension $d=3$, a non-vanishing bulk invariant $\Ch_{\{1,2,3\}}(A)$ implies that the surface states have a non-vanishing Hall conductance, which in a disordered situation is likely carried only by the states at zero energy \cite{PS}. Actually, generically there will be a flat band at zero energy in such three-dimensional chiral systems, with a surface density of states determined by the weak winding numbers of the bulk Fermi projection, see Theorem~0.0.3 in \cite{SSt}. 

\vspace{.2cm}

In general, it is not always clear whether there is a direct physical interpretation of the boundary invariant, in particular, if only weak bulk invariants are non-vanishing. One can, however, prove that in the presence of non-trivial bulk invariants there exist boun\-dary states that cannot be Anderson localized. The technical statement about this lack of Anderson localization is formulated using the eigenfunction correlator for the half-space operator $\widehat{H}$. For an open interval $\Delta$ with closure lying entirely in a bulk gap, it is defined by 
$$
\mathcal{Q}(n,m, \Delta) 
\;=\; 
\EE_\PM\;\sup_{f \in \Bb(\Delta)\,,\, \|f\|_\infty \leq 1}\; 
\big\|\langle n| f(\widehat{H})|m \rangle\big\|_2^2
\;,
$$
with the supremum taken over all bounded Borel functions $ \Bb(\Delta)$ supported by  $\Delta$ and  the Hilbert-Schmidt-norm on matrix degrees of freedom.  Then the Hamiltonian $\widehat{H}$ is said to satisfy strong dynamical localization in $\Delta$ if
\begin{equation}
\label{eq:correlator_bound_natural}
	\sup_{m\in \ZM^d}\; \sum_{n\in \ZM^d} \;\, (1+|n-m|)^{k} \, \mathcal{Q}(n,m,\Delta) 
	\; < \;\infty
\end{equation}
holds for each $k\in \NM$. It is a standard result of localization theory that \eqref{eq:correlator_bound_natural} holds whenever one can prove Anderson localization, either by the multi-scale or the fractional moment method. Now one can show (Section~5.5 in \cite{SSt}, with an earlier result requiring non-vanishing strong invariants in \cite{PS}) that, if $\Ch_{I}(P) \neq 0$ for some $I$ with $|I|>0$, then the bound \eqref{eq:correlator_bound_natural} cannot hold  for any $k\geq d-1$ and in any open interval $\Delta$ with closure lying in the bulk gap. Similarly, for a chiral Hamiltonian satisfying $\Ch_{I}(A) \neq 0$, the bound \eqref{eq:correlator_bound_natural} with $k\geq d-1$ cannot  hold in an interval $\Delta$ containing $0$, unless $0$ is an eigenvalue of $\widehat{H}$.

\vspace{.2cm}

\noindent {\bf Acknowledgements:} This work was supported by the DFG grant SCHU 1358/8-1.  


\newpage
.
\newpage

\begin{table}
\begin{center}

\begin{tabular}{|c|c|c|c||c||c|c|c|c|c|c|c|c|}
\hline
$j$ & $\!\!$TRS$\!\!$ & $\!\!$PHS$\!\!$ & $\!\!$CHS$\!\!$ & CAZ & $\!d=0,8\!$ & $\!d=1\!$ & $\!d=2\!$ & $d\!=3\!$ & $\!d=4\!$ & $\!d=5\!$ & $\!d=6\!$ & $\!d=7\!$
\\\hline\hline
$0$ & $0$ &$0$&$0$& A  & $\ZM$ &  & $\ZM$ &  & $\ZM$ &  & $\ZM$ &  
\\
$1$& $0$&$0$&$ 1$ & AIII & & $\ZM$ &  & $\ZM$  &  & $\ZM$ &  & $\ZM$
\\
\hline\hline
$0$ & $+1$&$0$&$0$ & AI &  $\ZM$ & &  & & $2 \, \ZM$ & & $\ZM_2$ & ${\ZM_2}$
\\
$1$ & $+1$&$+1$&$1$  & BDI & $\ZM_2$ &$\ZM$  & &  &  & $2 \, \ZM$ & & $\ZM_2$
\\
$2$ & $0$ &$+1$&$0$ & D & $\ZM_2$ & ${\ZM_2}$ & $\ZM$ &  & & & $2\,\ZM$ &
\\
$3$ & $-1$&$+1$&$1$  & DIII &  & $\ZM_2$  &  $\ZM_2$ &  $\ZM$ &  & & & $2\,\ZM$
\\
$4$ & $-1$&$0$&$0$ & AII & $2 \, \ZM$  & &  $\ZM_2$ & ${\ZM_2}$ & $\ZM$ & & &
\\
$5$ & $-1$&$-1$&$1$  & CII & & $2 \, \ZM$ &  & $\ZM_2$  & $\ZM_2$ & $\ZM$ & &
\\
$6$ & $0$ &$-1$&$0$ & C&  &  & $2\,\ZM$ &  & $\ZM_2$ & ${\ZM_2}$ & $\ZM$ &
\\
$7$ & $+1$&$-1$&$1$  &  CI &  & &   & $2 \, \ZM$ &  & $\ZM_2$ & $\ZM_2$ & $\ZM$
\\
[0.1cm]
\hline
\end{tabular}
\end{center}
\caption{{\it List of symmetry classes ordered by {\rm TRS}, {\rm PHS} and inherited {\rm CHS}. The signs correspond to $I^2=\pm\one$ and $K^2=\pm\one$ in \eqref{eq-SymClass}, the $0$ indicates that the symmetry is not present. Also the corresponding Cartan-Altland-Zirnbauer {\rm (CAZ)} label is listed. Then follow the strong invariants in dimension $d=0,\ldots,8$. }}
\label{tab-class}
\end{table}


\begin{thebibliography}{99}
\bibliographystyle{unsrt}


\bibitem{AMZ}
A.~Alldridge, C.~Max, M.~R.~Zirnbauer, {\sl Bulk-Boundary Correspondence for Disordered Free-Fermion Topological Phases}, Commun. Math. Phys. {\bf 377}, 1761-1821 (2020).



\bibitem{ASV} J.~C.~Avila, H.~Schulz-Baldes, C.~Villegas-Blas, {\sl Topological invariants of edge states for periodic two-dimensional models}, Math. Phys., Anal. Geom. {\bf 16}, 136-170 (2013). 

\bibitem{ASSS} J.~E.~Avron, L.~Sadun, J.~Segert, B.~Simon, {\sl Chern Numbers, Quaternions, and Berry's Phases in Fermi Systems}, Comm. Math. Phys. {\bf 124}, 595-727 (1989).

\bibitem{AS}
J. E. Avron, R. Seiler, {\sl Quantization of the Hall Conductance for Gerneral, Multiparticle Schr\"odinger Hamiltonians}, Phys. Rev. Lett. {\bf 54}, 259-262 (1985).

\bibitem{ASS0} J.~Avron,  R.~Seiler, B.~Simon, {\sl Homotopy and quantization in condensed matter physics}, Phys.
Rev. Lett. {\bf 51}, 51-53 (1983).

\bibitem{ASS} J.~Avron,  R.~Seiler, B.~Simon, {\sl The Charge deficiency, charge transport and comparison of dimensions}, Commun. Math. Phys. {\bf 159},  399-422 (1994).


\bibitem{Bel} J.~Bellissard, {\sl  K-theory of $C^*$-algebras in solid state physics}, in T.~Dorlas, M.~Hugenholtz, M.~Winnink, editors, Lecture Notes in Physics {\bf 257},  99-156, (Springer, Berlin, 1986).

\bibitem{BES} J.~Bellissard, A.~van Elst, H.~Schulz-Baldes, {\sl The non-commutative geometry of the quantum Hall effect}, J. Math. Phys. {\bf 35}, 5373 (1994).


\bibitem{BHZ} B.~A.~Bernevig, T.~L.~Hughes, S.-C.~Zhang, {\sl Quantum spin Hall effect and topological phase transition in HgTe quantum wells}, Science {\bf 314}, 1757-1761 (2006).

\bibitem{BCR} C.~Bourne, A.~L.~Carey, A.~Rennie, {\sl  The bulk-edge correspondence for the quantum Hall effect in Kasparov theory}, Lett. Math. Phys. {\bf 105}, 1253-1273 (2015).

\bibitem{BKR} C.~Bourne, J.~Kellendonk, A.~Rennie, {\sl The $K$-theoretic bulk-edge correspondence for topological insulators}, 
Ann. Henri Poincar\'e  {\bf18},  1833-1866  (2017).


\bibitem{CS} A.~Carey, H.~Schulz-Baldes, {\sl Spectral flow of monopole insertion in topological insulators}, Commun. Math. Phys. {\bf 370}, 895-923 (2019).

\bibitem{CKS} A.~Cerjan, L.~Koekenbier, H.~Schulz-Baldes, {\sl Spectral localizer for line-gapped non-hermitian systems}, J. Math. Phys. {\bf 64}, 082102 (2023).

\bibitem{Connes94} A.~Connes, {\sl Noncommutative geometry}, (Academic Press, San Diego, 1994).

\bibitem{DS0} G.~De~Nittis, H.~Schulz-Baldes, {\sl Spectral flows of dilations of Fredholm operators},  Canad. Math. Bull. {\bf 58}, 51-68  (2015).

\bibitem{DS} G.~De~Nittis, H.~Schulz-Baldes, {\sl Spectral flows associated to flux tubes}, Annales H. Poincare {\bf 17}, 1-35 (2016).


\bibitem{DS1} N.~Doll, H.~Schulz-Baldes, {\sl Approximate symmetries and conservation laws in topological insulators and associated $\ZM$-invariants}, Annals Phys.  {\bf 419}, 168238 (2020).

\bibitem{DS2} N.~Doll, H.~Schulz-Baldes, {\sl Skew localizer and $\ZM_2$-flows for real index pairings}, Advances Math. {\bf 392},  108038 (2021).

\bibitem{DSW} N.~Doll, H.~Schulz-Baldes, N.~Waterstraat, {\sl Spectral flow: A functional analytic and index-theoretic approach}, (De Gruyter, Berlin/Boston, 2023).



\bibitem{EG} P.~Elbau, G.-M.~Graf, {\sl Equality of bulk and edge Hall conductance revisited},  Commun. Math. Phys. {\bf 229}, 415-432 (2002). 

\bibitem{EGS} A.~Elgart, G.~M.~Graf, J.~H.~Schenker, {\sl Equality of the bulk and edge Hall conductances in a mobility gap},  Commun. Math. Phys. {\bf 259}, 185-221 (2005).




\bibitem{Gru} S.~Franca, A.~G.~Grushin, {\sl Obstructions in trivial metals as topological insulator zero-modes},
{\tt arXiv:2304.01983}.

\bibitem{GP} G.~M.~Graf, M.~Porta, {\sl Bulk-edge correspondence for two-dimensional topological insulators}, Commun.  Math. Phys. {\bf 324}, 851-895 (2013).


\bibitem{GS} J.~Grossmann, H.~Schulz-Baldes, {\sl Index pairings in presence of symmetries with applications to topological insulators}, Commun. Math. Phys.  {\bf 343}, 477-513 (2016).







\bibitem{Hal} F.~D.~M.~Haldane, {\sl Model for a Quantum Hall Effect without Landau Levels: Condensed-Matter Realization of the "Parity Anomaly''}, Phys. Rev. Lett. {\bf 61}, 2015 (1988).

\bibitem{Hat} Y.~Hatsugai, {\sl Chern number and edge states in the integer quantum Hall effect}, Phys. Rev. Lett. {\bf 71}, 3697-3700 (1993).

\bibitem{Hay} S.~Hayashi, {\sl An index theorem for quarter-plane Toeplitz operators via extended symbols and gapped Invariants related to corner states}, Commun. Math. Phys. {\bf 400}, 429-462 (2023).



\bibitem{KM2} C. L. Kane, E. J. Mele, {\sl Z(2) topological order and the quantum spin Hall effect}, Phys. Rev. Lett. {\bf 95}, 146802 (2005).


\bibitem{Kel} J.~Kellendonk,  {\sl Cyclic Cohomology for Graded $C^{*,r}$-algebras and Its Pairings with van Daele $K$-theory}, Commun. Math. Phys. {\bf 368}, 467-518 (2019).


\bibitem{KRS} J.~Kellendonk, T.~Richter, H.~Schulz-Baldes, {\sl Edge current channels and Chern numbers in the integer quantum Hall effect}, Rev. Math. Phys. {\bf 14}, 87-119 (2002).





\bibitem{KHW} M.~Kohmoto, B.~I.~Halperin, Y.-S.~Wu, {\sl Quantized hall effect in 3d periodic systems}, Physica B {\bf 184}, 30-33 (1993).


\bibitem{Kit0} A.~Kitaev, {\sl Anyons in an exactly solved model and beyond}, Annals of Phys. {\bf 321}, 2-111 (2006).

\bibitem{Kit} A.~Kitaev, {\sl Periodic table for topological insulators and superconductors}, (Advances in Theoretical Physics: Landau Memorial Conference) AIP Conference Proceedings {\bf 1134}, 22-30 (2009).


\bibitem{Kub} Y.~Kubota, {\sl The bulk–dislocation correspondence for weak topological insulators on screw–dislocated lattices}, J. Phys. A: Math. Theor {\bf 54}, 364001 (2021).

\bibitem{LP} B.~Leung, E.~Prodan, {\sl Bulk-boundary correspondence for topological insulators with quantized magneto-electric effect}, J. Phys. A: Math. Theo. {\bf 53}, 205203 (2020).




\bibitem{Lor} T.~A.~Loring, {\sl K-theory and pseudospectra for topological insulators}, Annals of Physics {\bf 356}, 383-416 (2015).


\bibitem{LH} T.~A.~Loring, M.~B.~Hastings,  {\sl Topological insulators and $C^*$-algebras: Theory and numerical practice}, Annals of Physics {\bf 326}, 1699-1759 (2011).

\bibitem{LS1} T.~A.~Loring, H. Schulz-Baldes, {\sl Finite volume calculation of $K$-theory invariants}, New York J. Math. {\bf 22}, 1111-1140 (2017).


\bibitem{LS2} T.~A.~Loring, H. Schulz-Baldes, {\sl The spectral localizer for even index pairings}, J. Noncommutative Geometry {\bf 14}, 1-23 (2020).

\bibitem{LSS} E.~Lozano Viesca, J.~Schober, H.~Schulz-Baldes, {\sl Chern numbers as half-signature of the spectral localizer},
J. Math. Phys. {\bf 60}, 072101 (2019).








\bibitem{PV} M.~Pimsner, D.~Voiculescu, {\sl Exact sequences for K-groups of certain cross-products of $C^*$ algebras}, J. Op. Theory {\bf 4}, 93-118 (1980).

\bibitem{Pro} E.~Prodan, {\sl Robustness of the spin-Chern number}, Phys. Rev. {\bf B 80}, 125327-125333 (2009).

\bibitem{Pro2} E.~Prodan, {\sl A Computational Non-Commutative Geometry Program for Disordered Topological Insulators}, Springer Briefs in Mathematical Physics, {\bf vol. 23}, (2017).

\bibitem{PLB} E.~Prodan, B.~Leung, J.~Bellissard, {\sl  The non-commutative $n$-th Chern number $(n\geq 0)$}, J. Phys. A: Math. Theor. {\bf 46}, 485202  (2013).

\bibitem{PS} E.~Prodan, H.~Schulz-Baldes, {\sl Bulk and Boundary Invariants for Complex Topological Insulators: From $K$-Theory to Physics}, (Springer International, Cham, 2016).

\bibitem{PS2} E.~Prodan, H.~Schulz-Baldes,  {\sl Generalized Connes-Chern characters in KK-theory with an 
application to weak invariants of topological insulators}, Rev. Math. Phys. {\bf 28}, 1650024 (2016).

\bibitem{QHZ} X.-L.~Qi, T.~L. Hughes, S.-C.~Zhang, {\sl Topological field theory of time-reversal invariant insulators}, Phys. Rev. B {\bf78}, 195424 (2008).


\bibitem{RSFL}
S.~Ryu, A.~P. Schnyder, A.~Furusaki,  A.~W.~W. Ludwig, {\sl  Topological insulators and superconductors: tenfold way and  dimensional hierarchy}, New J. Phys. {\bf 12}, 065010 (2010).

\bibitem{SaS} C.~Sadel, H.~Schulz-Baldes, {\sl Topological boundary invariants for Floquet systems and quantum walks},
Math. Phys., Anal. Geom. {\bf 20}, 22 (2017).

\bibitem{SB1} H.~Schulz-Baldes, {\sl Persistence of spin edge currents in disordered quantum spin Hall systems},  Commun. Math. Phys. {\bf 324},  589-600 (2013).

\bibitem{SB5} H.~Schulz-Baldes, {\sl $\ZM_2$-indices and factorization properties of odd symmetric Fredholm operators}, Documenta Math. {\bf 20}, 1481-1500 (2015).

\bibitem{SB6} H.~Schulz-Baldes, {\sl Topological insulators from the perspective of non-commutative geometry and index theory},
Jahresbericht der deutschen Mathematiker-Vereinigung {\bf 118},  247-273 (2016).




\bibitem{SKR} H.~Schulz-Baldes, J.~Kellendonk, T.~Richter, {\sl Simultaneous quantization of edge and bulk Hall conductivity}, J.  Phys. A: Math. Gen. {\bf 33}, L27 (2000).

\bibitem{SS1} H. Schulz-Baldes, T.~Stoiber, {\sl The spectral localizer for semifinite spectral triples},  Proc. AMS {\bf 149}, 121-134 (2021).

\bibitem{SS2} H.~Schulz-Baldes, T.~Stoiber {\sl Invariants of disordered semimetals via the spectral localizer}, Europhys. Letters {\bf 136}, 27001 (2021).

\bibitem{SSt} H.~Schulz-Baldes, T.~Stoiber, {\sl Harmonic analysis in operator algebras and its applications to index theory and topological solid state systems}, Springer Series Mathematical Physics Studies, 220 pages (Springer Int. Pub., Cham, Switzerland, 2022)



\bibitem{STo} H.~Schulz-Baldes, D.~Toniolo, {\sl Dimensional reduction and scattering formulation for even topological invariants}, Commun. Math. Phys. {\bf 381}, 119-142 (2021).

\bibitem{ShT} J.~Shapiro, C.~Tauber,  {\sl Strongly disordered Floquet topological systems}, Annales H. Poincar\'e {\bf 20}, 1837-1875 (2019).



\bibitem{SCR} M.~Stone, C.-K,~Chiu, A.~Roy, {\sl Symmetries, dimensions and topological insulators: the mechanism behind the face of the Bott clock},  J. Phys. A: Math. Theor. {\bf 44}, 045001 (2011).





\bibitem{Thi} G.~C.~Thiang, {\sl On the K-theoretic classification of topological phases of matter}, Annales H. Poincar\'e {\bf 17},  757-794 (2016). 

\bibitem{TKN} D. J. Thouless, M. Kohmoto, M. P. Nightingale, M. den Nijs, {\sl Quantized Hall Coinductance in a Two-Dimensional Periodic Potential}, Phys. Rev. Lett. {\bf 49}, 405-408 (1982).




\end{thebibliography}
\end{document}